# Mobile Academy: A Ubiquitous Mobile Learning (mLearning) Platform

Mahdi H. Miraz[1,2], Sajid Khan[3], Moniruzzaman Bhuiyan[4] and Peter Excell[1]

[1]Department of Computing, Glyndŵr University, Wrexham, UK
`(m.miraz||p.excell)@glyndwr.ac.uk`

[2]Department of Computer Science & Software Engineering, University of Hail, KSA
`m.miraz@uoh.edu.sa`

[3]Samsung Research & Development Institute, Dhaka, Bangladesh
`sajid.khan@samsung.com`

[4]Institute of Information Technology, University of Dhaka, Bangladesh
`mb@du.ac.bd`

*Abstract*- **The paper reports on an ongoing research project into the development of "Mobile Academy", an Android-based mobile learning (mLearning) application (app). The project comprises three major phases: requirement analysis, application development and testing and evaluation. To satisfy the user requirement analysis, a detailed ethnographic study was conducted to investigate how people from different background use mobile devices for learning purposes. The initial analysis and evaluation of the first version of the projected app demonstrates very promising results. Making use of the app seemed to have, in general, a positive dimension in facilitating educational use of mobile devices.**

## I. INTRODUCTION

Despite the predictions of some sceptics, the growth of computational power continues in broad accordance with Moore's law [1,2,3]. On the other hand, the prices of computing and networking equipment per unit performance metric (e.g. MIPs or Mbit/s), including mobile devices and charges to access the Internet, are decreasing at an inverse rate. Hence, the usage of the Internet almost anywhere in the developed (and increasingly in the developing) World has become a norm. Further, the widespread adoption of smartphones and, more recently, the concept of the Internet of Everything (IoE) has led to the inspiration of using mobile devices for any kind of internet use, including even financial transactions. Such widespread mobile usage of digital and electronic techniques, technologies and applications holds huge promise to widen the horizon of teaching and learning, especially through mLearning.

The goal of the present research is to develop a ubiquitous mobile learning platform for universal types of users to facilitate teaching and learning on the move. Previous ethnographic surveys conducted by the authors partially fulfill the requirements analysis and suggest the development of an mLearning platform to be simultaneously used, in particular, by people from different cross-national boundaries: this cross-national (or cross-cultural) aspect is believed to distinguish the work from other examples currently known.

The initial phase of the proposed project, developing the first iteration of the Mobile Academy app for Android-based handheld devices, has been implemented, tested and verified to demonstrate the merits and capabilities of the scheme through a set of experiments. However, the investigation will continue to improve for the app to be able to adopt a wider range of real world usage.

## II. RESEARCH BACKGROUND

Mobile learning is seen as one of the leading edge teaching and learning technologies [4]. However, there has been no formal definition of mobile learning so far and hence the perception of it varies among individuals.

However, the definition as outlined by Sharples *et al.* can be considered as a working one. According to them, mobile learning is considered as the "process of coming to know through conversations across multiple contexts among people and personal interactive technologies" [5]. The supportive technologies here include any form of handheld devices that can support learning and teaching, such as smart phones, personal digital assistants (PDAs), tablets or even a simple mobile phone. Although it is obvious that laptops are somewhat mobile, they are excluded from the list [6].

Due to the pervasive adoption [7] of popular internet and networking modalities such as Social Media, Social Networking, Mobile instant messaging and the like, as part of a continuous development process, universities and other higher education providers are required to become accustomed to, and to adopt them to facilitate learning and education. Mobile communication, simultaneously with other Internet communication technologies, is becoming wide-spread as a means of education and is expected to bridge the gap [8] between formal and informal learning and teaching methodologies. The stakeholders now have to pay attention to how people embrace and live with the new technologies [9], as this trend will greatly contribute to the dramatic transformation of education systems' characteristics and traits.

The multidimensional and exponentially increasing use of mobile technology is influencing cultural practice and facilitates novel contexts for learning [10], although the integration of mobile technologies in teaching is observing a little slower rate than social media, due to the fact that the instructors themselves first need to be equipped with the knowledge of how to use them [11]. However, from the way that mobile devices and networking technologies are becoming a routine part of daily life, it can be foreseen that mLearning will soon be widely adopted by the education sectors around the globe.

However, like any other technologies, mobile phones and other handheld devices suffer from technical limitations which should be carefully considered. These limitations have been categorised into three major groups [12] based on users' pedagogical, psychological and technical limitations. The aim of the present project is to develop a ubiquitous mobile learning app, "Mobile Academy", to address these limitations.

III. RESEARCH METHODOLOGY AND DESIGN

The complete project has been divided into three phases as follows: 1. The Requirements Analysis, 2. Design and Development and 3. Testing and Evaluation.

The Requirements analysis phase involves identifying the needs of the users: the students as well as the teachers. Opinions from both the parties were sought and an extensive survey of existing mLearning apps was conducted. Thus, based on the needs identified, strategies for developing the required functionalities to satisfy the requirements for the high-school case were developed, as described in Table-1; use case diagrams, as shown in Fig. 1, were also used.

The second phase of the project involves designing and developing the app: some coding was also involved at this stage. The overall functionalities and navigation while using the app can be best described using the following flowchart, as shown in Fig. 2.

The app was developed using Android 4.2 (Jelly Bean). It is thus compatible with any handheld devices running Android 2.3 (Gingerbread) or above. User Centric Design (UCD), Participatory Design and other design and development methods of Human-Computer Interaction (HCI) were deployed at this stage.

TABLE I
FUNCTIONALITY REQUIREMENT ANALYSIS

| No | Requisite | Applied Event/Function |
|---|---|---|
| 1. | Gmail Login | Opening the application |
| 2. | School Creation/Editing/Delete | Teacher account |
| 3. | All School Lists | Schools |
| 4. | Settings for an examination | Exam name |
| 5. | Uploading documents | Teacher account chapter list upload |
| 6. | Viewing documents | Schools chapter list |
| 7. | Viewing all results | Results |
| 8. | Terminating interface | Exit |

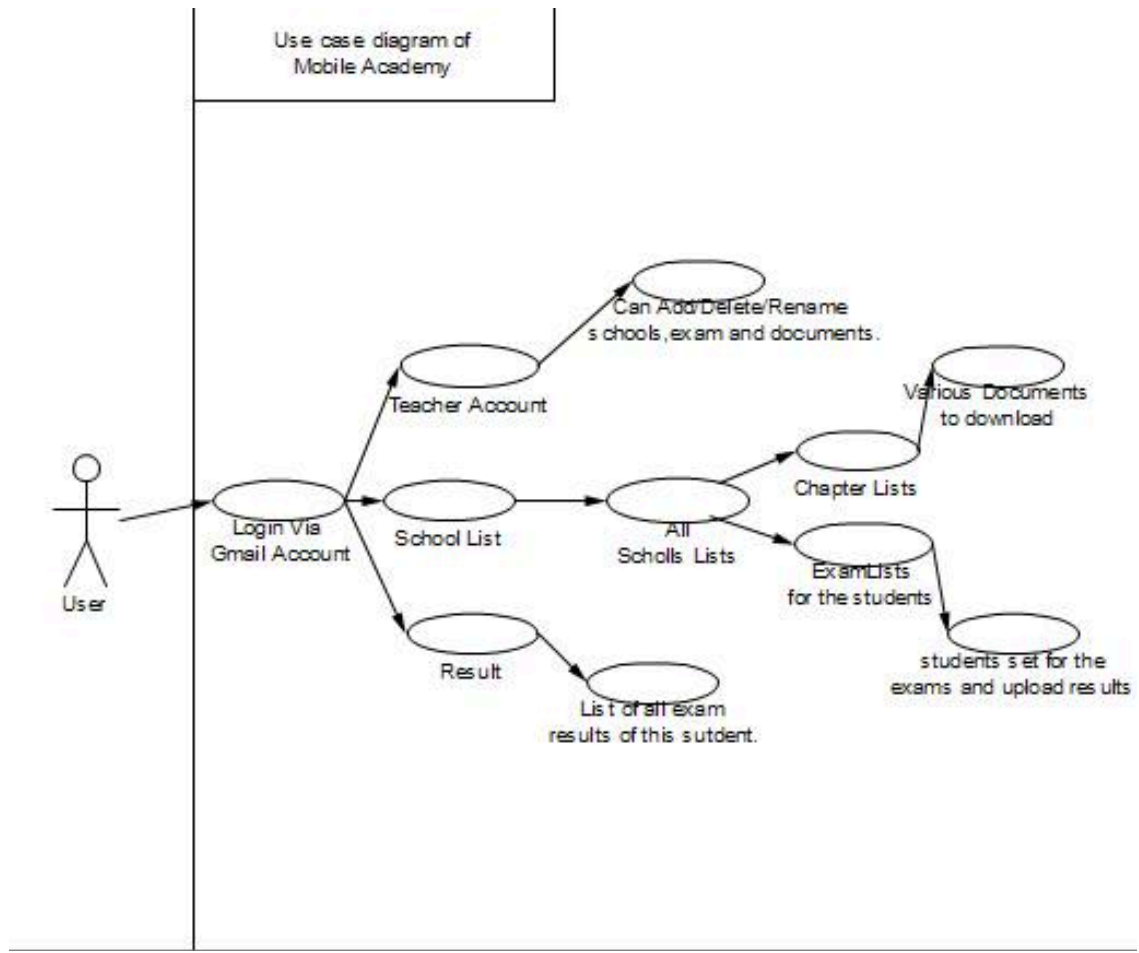

Figure 1. Use Case Diagram for the first iteration of the Mobile Academy app

As an example, the creation of a new school instance within the app, using a teacher's account is shown in Fig. 3: Fig. 4 demonstrates the process of setting up an exam. The teachers can set the questions and the answers so that, upon completion, the results are immediately calculated by the system and displayed. The exam can be protected by instructor set keys (password) and can be validated for a specific time only. The grades are saved into the students' records. Fig. 5 displays the options for the students regarding enrolling into different schools: this can be restricted by instructor set enrolment keys (password). However, this process of restricting the enrolment is optional. The instructors can opt out if they so choose.

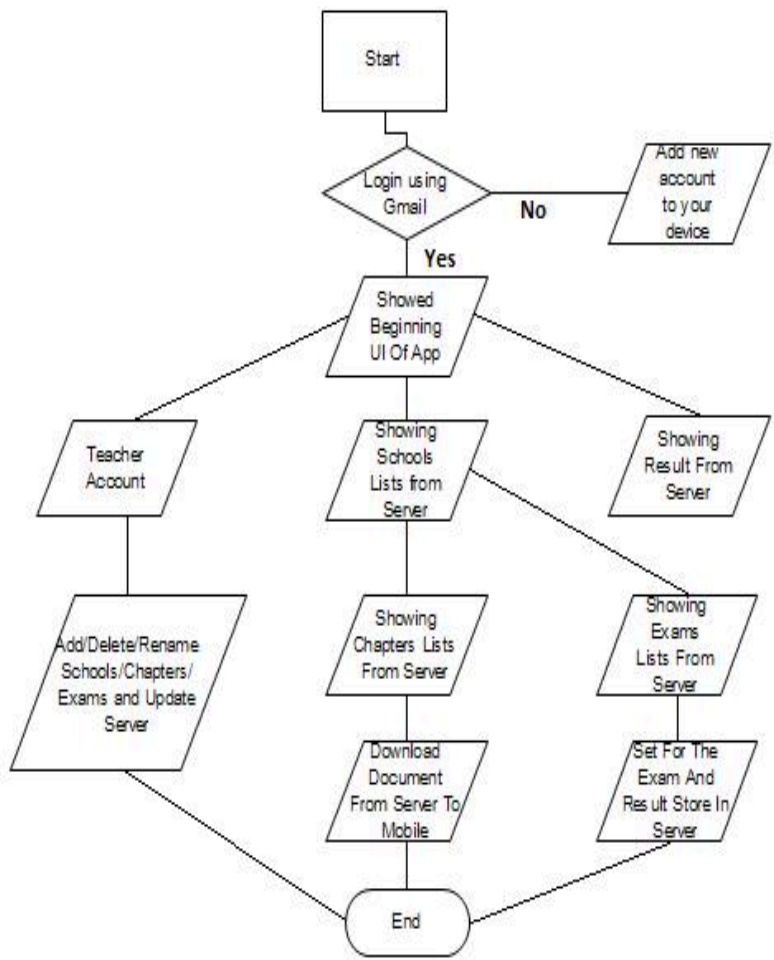

Figure 2. Functionality and navigation flowchart of the Mobile Academy app

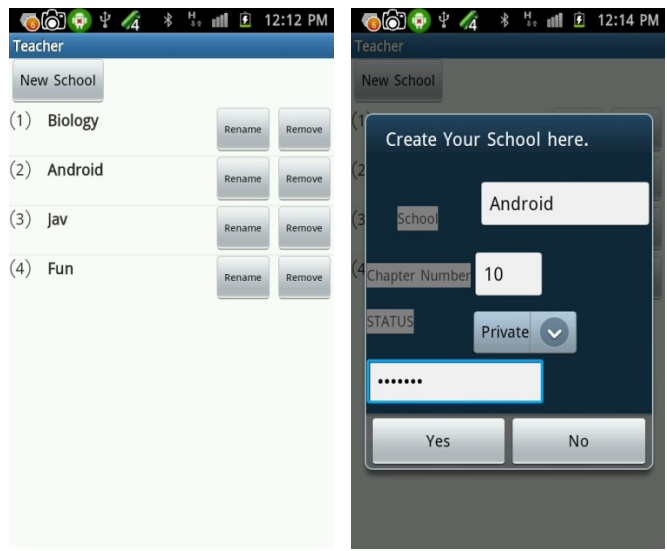

Figure 3. Creation of new school instance within the app, using a teacher's account.

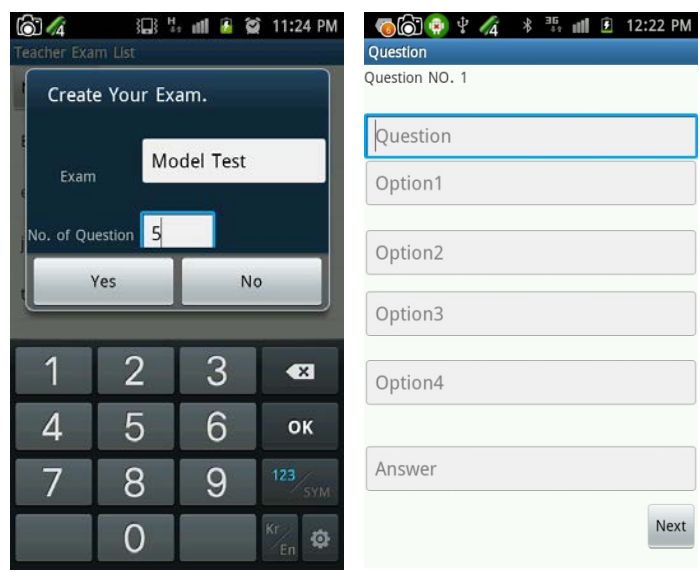

Figure 4. Exam setting up process.

### III. TESTING AND EVALUATION

As the app was developed using Android 4.2 (Jelly Bean), it is thus expected to be compatible with any handheld devices running Android 2.3 (Gingerbread) or above. However, this does not guarantee full compatibility. Hence, to confirm such compatibility, cross-device tests were conducted and satisfactory results were obtained. Some devices, especially smartphones with very much smaller screens, suffered some usability problem at the first iteration of the app. This was then solved by modifying the overall design of the app. Fig. 6 demonstrates an example of playing, while conducting the cross-device compatibility tests, video lecture uploaded by the teacher. It can also play animations and audio files. The functionalities of the app were thoroughly tested and any other bugs found were resolved.

Proc. of the Int. Conf. on eBusiness, eCommerce, eManagement, eLearning and eGovernance (IC5E 2014)

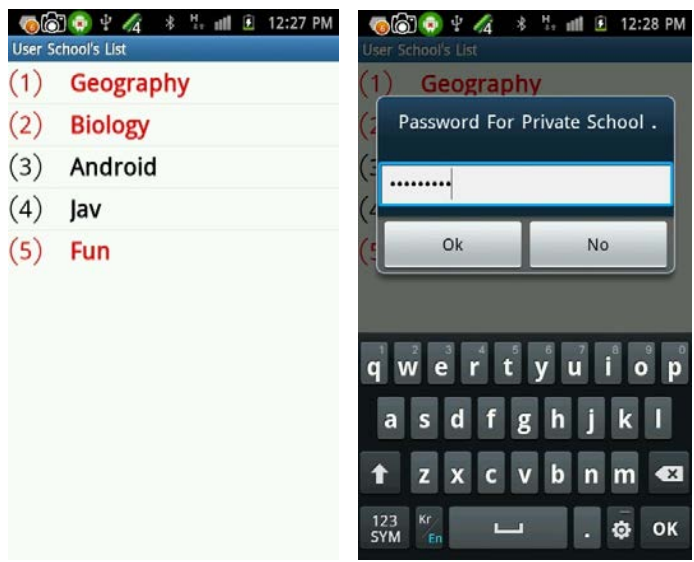

Figure 5. Students' options for enrolling in different schools.

The app was informally used for testing purposes while teaching, at the University of Hail (Saudi Arabia), for the Data and Computer Communication (COE 341) and Multimedia Systems (SWE 423) courses, consisting of a total of 16 students. All of the students were already familiar with using at least one e/m-learning app such as Edmodo, Khan Academy, Cisco NetSpace and similar. Due to time constraints, a detailed evaluation had not yet been conducted at the time of writing this paper. However, most of the students were satisfied with the functionality of the app. Features such as attending exams, reviewing grades and availability of the study materials at any place and any time were identified as the most popular ones. Some of the participants suggested implementing an online version of the Mobile Academy and integrating it with the app.

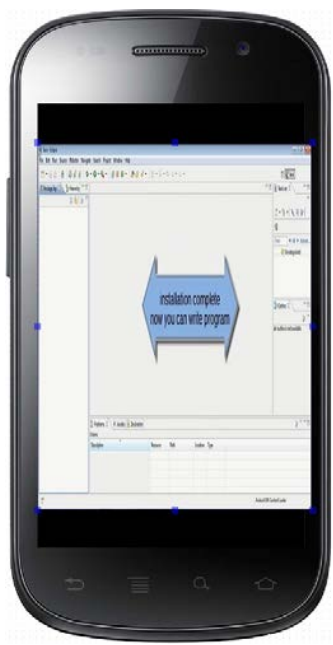

Figure 5. Playing Video Lecture.

IV. CONCLUSION

The project involved designing, developing and testing an mLearning app to be used by a wide range of users from different backgrounds. The initial version of the app has now been developed and tested for proper functionalities as well as cross device compatibility. A small scale initial usability test was conducted which provided positive results. However, the app is planned to undergo more iterations and future large scale usability tests. Cross-cultural usability tests are also required to achieve universal usability.